\documentclass[useAMS,usenatbib,fleqn]{mnras}
\pdfoutput=1
\usepackage{newtxtext,newtxmath}
\usepackage{algorithm,algorithmic}
\usepackage[T1]{fontenc}
\usepackage{ae,aecompl}

\usepackage[dvips]{graphicx}
\usepackage{epsfig,amssymb,amsmath,color}
\newcommand{\beq}{\begin{equation}}
\newcommand{\eeq}{\end{equation}}
\newcommand{\beqn}{\begin{eqnarray}}
\newcommand{\eeqn}{\end{eqnarray}}

\def\bmath#1{\mbox{\boldmath$#1$}}

\long\def\symbolfootnote[#1]#2{\begingroup%
\def\thefootnote{\fnsymbol{footnote}}\footnote[#1]{#2}\endgroup}

\usepackage{pstricks,pst-text}
\floatname{algorithm}{Algorithm}

\title[Stochastic Calibration]{Stochastic Calibration of Radio Interferometers}
\author[Yatawatta]{Sarod Yatawatta$^{1}$\thanks{E-mail:
yatawatta@astron.nl}\\
$^{1}$ASTRON, Oude Hoogeveensedijk 4, 7991 PD Dwingeloo, The Netherlands}

\begin{document}
\date{\today}
\pagerange{\pageref{firstpage}--\pageref{lastpage}} \pubyear{2018}
\maketitle
\label{firstpage}
\begin{abstract}
With ever increasing data rates produced by modern radio telescopes like LOFAR and future telescopes like the SKA, many data processing steps are overwhelmed by the amount of data that needs to be handled using limited compute resources. Calibration is one such operation that dominates the overall data processing computational cost, nonetheless, it is an essential operation to reach many science goals. Calibration algorithms do exist that scale well with the number of stations of an array and the number of directions being calibrated. However, the remaining bottleneck is the raw data volume, which scales with the number of baselines, and which is proportional to the square of the number of stations. We propose a 'stochastic' calibration strategy where we only read in a mini-batch of data for obtaining calibration solutions, as opposed to reading the full batch of data being calibrated. {Nonetheless, we obtain solutions that are valid for the full batch of data.} Normally, data need to be averaged before calibration is performed to accommodate the data in size-limited compute memory. Stochastic calibration overcomes the need for data averaging before any calibration can be performed, and offers many advantages including: enabling the mitigation of faint radio frequency interference; better removal of strong celestial sources from the data; and better detection and spatial localization of fast radio transients.
\end{abstract}
\section{Introduction}
The science goals of modern radio telescopes are diverse and require the highest quality data to be delivered to the end users. In order to achieve this, data are taken at the highest resolution in time and in frequency, so that radio frequency interference (RFI) mitigation  \citep{Wilensky2019} can be carried out satisfactorily. An equally important data processing step is the elimination of systematic errors from the data, also called as {\em calibration}. Systematic errors are introduced by the Earth atmosphere and by the instrument itself. Calibration is essentially an estimation problem and for its success the  data need to have sufficient signal to noise ratio (SNR), in other words, sufficient number of data samples need to be considered together (thus increasing the effective SNR). However, with limited compute memory, the number of  data samples that can be accommodated in compute memory is limited. Hence, the commonly used practice is to average the data before any calibration is performed. On the one hand, this is inevitable due to limited memory but on the other hand, averaging also loses some valuable information.

\begin{figure*}
\begin{minipage}{0.98\linewidth}
\begin{center}
\centering
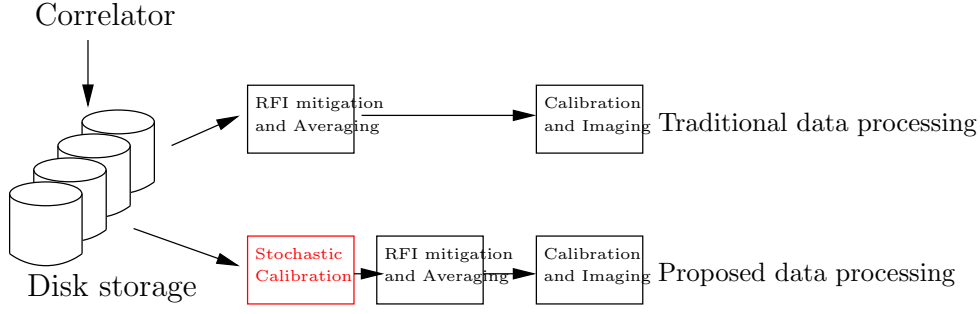
\vspace{0.1in}
\end{center}
\end{minipage}
\caption{The proposed calibration of data at highest resolution, compared with traditional data processing.\label{new_cal}}
\end{figure*}

In this paper, we propose a paradigm shift in the processing of radio interferometric data as shown in Fig. \ref{new_cal}. We re-use a widely used concept in modern machine learning -- i.e., stochastic learning or training, and introduce {\em stochastic} calibration of radio interferometric data. We calibrate data at the highest resolution -- i.e., at the same resolution where RFI mitigation is performed. Normally, calibrating data at this resolution would require a huge amount of compute memory. We overcome that by working with a subset of data at each iteration of calibration, we call this subset of data a {\em mini-batch}. These mini-batches are sequentially fed to the calibration algorithm, and at convergence, we find a solution that is valid for the full dataset being calibrated. Note that what we call as the {\em full dataset} here is the dataset that is within a specific time and frequency interval, in other words, the domain of the calibration solutions is defined by this time and frequency interval. Moreover, this full dataset is still only a fraction of the total data being calibrated in a long observation where multiple calibration solutions are obtained.

Working with mini-batches of data introduces a fundamental problem -- increased variance \citep{VarReduc} due to the lower SNR compared with the full dataset. This has already been noticed in calibration as well, for instance by \cite{Jeffs06} in their demixed peeling approach of calibration. There are many ways of reducing this variance and we refer the reader to e.g. \citep{robbins1951,VarReduc,Adam} for some widely used methods in machine learning. There is however a subtle difference in most machine learning problems and calibration -- i.e., the size of the data. In most machine learning problems, the full dataset can be pre-loaded into compute memory but this is nearly impossible for radio interferometric data at the highest resolution. Hence the data need to be read from disk during each iteration of calibration. The number of times the full dataset (divided into many mini-batches) is passed through the learning algorithm is called as an {\em epoch} in machine learning jargon and we use the same term here as well. Because we read the data from disk storage during calibration, it is also important minimize the number of epochs needed for finding a satisfactory solution. With this objective, we have already introduced a stochastic, limited memory Broyden Fletcher Goldfarb Shanno (LBFGS) algorithm in \citep{escience2018,DSW2019}. Compared with the commonly used gradient descent based algorithms \citep{robbins1951,Adam} in machine learning, the LBFGS algorithm has faster convergence \cite{Fletcher,Liu1989} and hence need lower number of epochs in calibration as we show later.

There is a wide variety of calibration algorithms (too numerous to list here, see e.g. \cite{DSW2019} and references therein), each having their own merits and demerits. However, one common aspect of these algorithms is that they all operate in full batch mode. By introducing stochastic calibration and enabling calibration of data at the highest resolution, we get many additional benefits as we describe below.

\begin{itemize}
\item RFI mitigation will work better if the signals from the celestial sources are subtracted from the data \citep{Wilensky2019}. Hence, the residual of stochastic calibration (where the sky is subtracted) will reveal many more weak RFI signals. { One caveat here is that the RFI mitigation is dependent on the accuracy (and completeness) of the sky model used in calibration.}
\item The spatial localization of fast radio bursts (FRB) \citep{Chat2017} need calibrated data at the highest resolution and stochastic calibration is an obvious choice to provide this.
\item The removal of strong celestial sources (the Sun, Cassiopeia A, Cygnus A etc.) that appear far away from the field of view is best done at the highest resolution of data. Stochastic calibration is an improvement to \citep{Jeffs06} in this regard.
\item Radio polarimetric science (rotation measure synthesis) \cite{Brentjens2005,Schnitz2015} can also benefit from stochastic calibration. First, by preserving the frequency resolution of the data, we can maximize the range of Faraday depth in the data. Secondly, by preserving the time resolution during calibration, we can overcome depolarization due to rapid Faraday rotation of the data.
\item Low power devices provide an energy efficient alternative for processing of data by telescopes such as the SKA \citep{SKA0,Spreeuw2019}. However, such devices have limited compute memory and stochastic calibration provides a feasible calibration algorithm in that case. 
\item Distributed calibration where the data is calibrated using a network of compute agents { that exchange information available at multiple frequencies} is shown to give better results \citep{DCAL,Brossard2016,DMUX,OLLIER2018} { than conventional single-frequency calibration}. For instance, in \citep{Patil2017}, about $300$ sub-bands are calibrated using about $60$ compute agents. The data used in this case is averaged by a factor of about $60$ in frequency before calibration is performed. If the same data are calibrated at the original resolution, the number of compute agents needed would increase from $300$ to about $18000$ which would overwhelm the network. We propose a distributed stochastic calibration scheme for this situation where the number of compute agents that need to communicate is minimized.
\item The calibration for the instrumental pass band (bandpass calibration) is normally done using large time intervals because the bandpass is assumed to vary very slowly with time. With stochastic calibration, we can use large time intervals for bandpass calibration.
\end{itemize}

We emphasize that as shown in Fig. \ref{new_cal}, stochastic calibration is not a replacement for traditional calibration that is done at a later stage of data processing. On the contrary, it is an enhancement to traditional data processing where all existing data processing stages can follow stochastic calibration.

The rest of the paper is organized is as follows. In section \ref{sec:model}, we introduce the data model we use in radio interferometry. In section \ref{sec:calib}, we present distributed stochastic calibration of multi-frequency radio interferometric data. In section \ref{sec:simul}, we  compare the performance of the stochastic LBFGS algorithm \cite{DSW2019} to commonly used first order learning algorithms used in machine learning \cite{Adam} in stochastic calibration using PyTorch \cite{paszke}. We also provide an example of distributed stochastic calibration in section \ref{sec:simul}. Finally, we draw our conclusions in section \ref{sec:conc}.

{\em Notation}: Lower case bold letters refer to column vectors (e.g. ${\bf y}$). Upper case bold letters refer to matrices (e.g. ${\bf C}$). Unless otherwise stated, all parameters are complex numbers. The set of complex numbers is given as ${\mathbb C}$ and the set of real numbers as  ${\mathbb R}$. The matrix inverse, pseudo-inverse, transpose, Hermitian transpose, and conjugation are referred to as $(\cdot)^{-1}$, $(\cdot)^{\dagger}$, $(\cdot)^{T}$, $(\cdot)^{H}$, $(\cdot)^{\star}$, respectively. The matrix Kronecker product is given by $\otimes$. The vectorized representation of a matrix is given by $\mathrm{vec}(\cdot)$. The $i$-th element of a vector ${\bf y}$ is given by $y[i]$. The identity matrix of size $N$ is given by ${\bf I}_N$. All logarithms are to the base $e$, unless stated otherwise. The Frobenius norm is given by $\|\cdot \|$. Rounding up to the nearest integer is done by $\lceil \cdot \rceil$.

\section{Radio interferometric data model}\label{sec:model}
In this section, we give an overview of the data model we use, especially in relation to stochastic calibration. The interferometric signal formed by the receiver pair $p$ and $q$ is given as \citep{HBS}
\beq \label{V}
{\bf V}_{pq}=\sum_{i=1}^K {\bf J}_{pi} {\bf C}_{pqi} {\bf J}_{qi}^H + {\bf N}_{pq}
\eeq
where we have signals from $K$ directions in the sky being received. The systematic errors along the $i$-th direction are given by ${\bf J}_{pi}$ and ${\bf J}_{qi}$ ($\in {\mathbb C}^{2\times 2}$) for the $p$-th and $q$-th station, respectively. The source { coherency} ${\bf C}_{pqi}$ ($\in {\mathbb C}^{2\times 2}$) is generally well known and calculated using a sky model \citep{TMS}. The noise ${\bf N}_{pq}$  ($\in {\mathbb C}^{2\times 2}$) is assumed to have complex circular Gaussian entries. All entries of (\ref{V}) are time and frequency dependent and this is implicitly assumed throughout the paper. { The total number of stations is $N$, therefore $p,q\in[1,N]$ and the total number of baselines per given time and frequency is $N(N-1)/2$.}

Calibration is the determination of ${\bf J}_{pi}$ and ${\bf J}_{qi}$ in (\ref{V}) for all $p,q,i$ and for the full time and frequency domain of the data. Since there are too many time and frequency points at which data are taken, solutions for ${\bf J}_{pi}$ and ${\bf J}_{qi}$ are obtained for finite time and frequency intervals, that cover many data points. In order to use our stochastic LBFGS \citep{DSW2019} algorithm for calibration, we need to convert (\ref{V}) to a model with real values. First, we vectorize (\ref{V}) and get
\beq \label{vecV}
{\bf v}_{pq}={\bf s}_{pq}({\bmath \theta}) +{\bf n}_{pq}
\eeq
where ${\bf s}_{pq}({\bmath \theta})=\sum_{i=1}^{K}({\bf J}_{qi}^{\star}\otimes{\bf J}_{pi}) vec({\bf C}_{pqi})$, ${\bf v}_{pq}=vec({\bf V}_{pq})$, and ${\bf n}_{pq}=vec({\bf N}_{pq})$. We represent the parameters ${\bf J}_{pi}$ and ${\bf J}_{qi}$ (that are ${\mathbb C}^{2\times 2}$ matrices) as ${\bmath \theta}$, a vector of real parameters of length $8NK$ ($\in {\mathbb R}^{8NK\times 1}${, the factor $8$ comes from representing the $4$ complex cross-correlations as real values}). Consider that there are $T$ samples in time and $B$ samples in frequency within the time and frequency interval where calibration is performed. For a full observation, there will be many such intervals to cover the full integration time and bandwidth. We stack all data points within the calibration interval into vectors as
\beqn \label{modvec}
{\bf x}=[\mathrm{real}({\bf v}_{12}^T),\mathrm{imag}({\bf v}_{12}^T),\mathrm{real}({\bf v}_{13}^T),\ldots]^T\\\nonumber
{ {\bf m}({\bmath \theta})=[\mathrm{real}\left({\bf s({\bmath \theta})}_{12}^T\right),\mathrm{imag}\left({\bf s({\bmath \theta})}_{12}^T\right),\mathrm{real}\left({\bf s({\bmath \theta})}_{13}^T\right),\ldots]^T}\\\nonumber
\eeqn
where ${\bf x}$ and  ${\bf m}({\bmath \theta})$ are vectors of size $N(N-1)/2\times 8\times T \times B$ ($\in  {\mathbb R}^{4TBN(N-1)}$). { In (\ref{modvec}), ${\bf x}$ represents the data being calibrated and  ${\bf m}({\bmath \theta})$ represents the predicted model visibilities based on the current value of ${\bmath \theta}$.}

{ We use a robust noise model for ${\bf n}_{pq}$ as in \cite{Kaz3} during calibration. For maximum likelihood estimation, the negative log-likelihood of the data is minimized. Ignoring the terms independent of ${\bmath \theta}$, the cost function to be minimized becomes}
\beq \label{gLBFGS}
g({\bmath \theta})=\sum_{i=1}^{4TBN(N-1)} \log\left(1 +\frac{\left({\bf x}[i]-{\bf m}({\bmath \theta})[i]\right)^2}{\nu}\right)
\eeq
where ${\bf x}[i]$ and ${\bf m}({\bmath \theta})[i]$ represent the $i$-th elements of ${\bf x}$ and ${\bf m}({\bmath \theta})$, respectively, and $\nu$ is the degrees of freedom \citep{Kaz3}. { It is possible to select the most suitable value for $\nu$ based on the data itself as in \citep{Kaz3}. However, as we are also after reducing the computational cost of calibration, we select a low value $\nu=2$ for improved robustness (note that by making $\nu \rightarrow \infty$, we get a Gaussian noise model).}

In (\ref{gLBFGS}), the total number of data points used to evaluate the cost function is $4TBN(N-1)$, and this is the full batch size. In time samples, the full batch size is $T$. In stochastic calibration, we use $M$ time samples $1\le M \le T$ to obtain a solution for (\ref{gLBFGS}). Therefore, the full batch is divided into $\lceil T/M \rceil$ mini-batches. In other words, if the $i$-th mini-batch has $g_i({\bmath \theta})$ as the cost function, 
\beq \label{minicost}
g({\bmath \theta})=\sum_{i=1}^{\lceil T/M \rceil} g_i({\bmath \theta}).
\eeq
Note also that in (\ref{minicost}), in spite of working with mini-batches of data, we still find one solution for ${\bmath \theta}$ that minimizes $g({\bmath \theta})$, covering the full $4TBN(N-1)$ data points. Therefore, the time and frequency domain of the solution for ${\bmath \theta}$ is determined by the full batch of data. Because we work with $M$ time slots instead of $T$, we require less memory and computation if $M\ll T$. The main drawback of this approach however is increased variance \citep{robbins1951} and minimizing this has been well studied (e.g., \cite{VarReduc}). The stochastic LBFGS algorithm we have developed in \citep{DSW2019} also takes into account the increased variance as are other versions of stochastic LBFGS \citep{Berahas,Bolla,Li2018}. In section \ref{sec:calib}, we enhance the performance of stochastic calibration by exploiting the continuity of ${\bmath \theta}$ over frequency and using the full bandwidth of the observation into our advantage as in \cite{DCAL}.

\section{Distributed stochastic calibration}\label{sec:calib}
\begin{figure*}
\begin{minipage}{0.98\linewidth}
\begin{center}
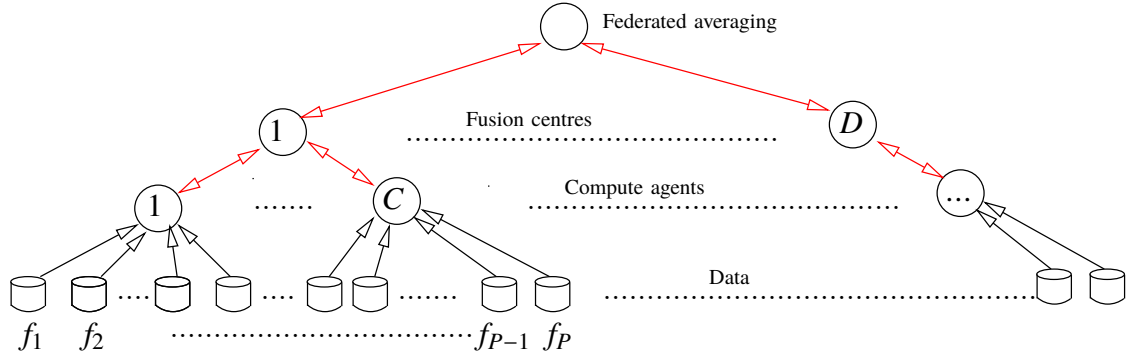\\
\end{center}
\end{minipage}
\caption{Distributed stochastic calibration framework. Data are distributed across multiple networks. There are $D$ fusion centres and the $1$-st fusion center is connected to $C$ compute agents that access the data stored local to them. The total number of datasets (frequencies) accessed by compute agents connected to the $1$-st fusion centre is $P$. The frequencies of the data handled by these $P$ compute agents are given by  $f_1,f_2,\ldots,f_P$  respectively. The $D$ fusion centres are connected to a higher level fusion centre where only averaging is performed.\label{block}}
\end{figure*}

Calibration is performed over a small time and frequency interval compared with the full observation that has a large integration time and a wide bandwidth. We propose a scheme where we can work with mini-batches of data and at the same time, exploit the continuity of systematic errors over frequency as in \citep{DCAL,DMUX}. We introduce the distributed computing framework as shown in Fig. \ref{block}, which is a refinement of our previous work (e.g. Fig. 1 of \cite{DMUX}). The main difference in Fig. \ref{block} compared to our previous work is that we have $D$ fusion centres instead of just one. On top of this, we have a higher level centre where only averaging is performed. 

The motivation behind the framework shown in Fig. \ref{block} is to handle significantly more frequency channels than in our previous work \citep{DMUX}. As we discussed before (and as shown in Fig. \ref{new_cal}), data at the highest resolution will have orders of magnitudes more channels than data that is averaged. Therefore, adopting a strategy with only one fusion centre as in \citep{DMUX} would be prohibitive in terms of the network bandwidth required at the fusion centre. In order to overcome this, we propose a hierarchy, where we have $D$ fusion centres $D>1$ and each fusion centre is connected to a subset of compute agents. In Fig. \ref{block} for instance, the $1$-st fusion centre is connected to $C$ compute agents and these $C$ compute agents have access to data at $P$ frequencies. If we have a similar data distribution in other fusion centres as well, the top level federated averaging centre gets information only from $D$ agents instead of $PD$ (or $CD$ with multiplexing) as in our previous schemes.

Each fusion centre and the compute agents connected to it perform consensus optimization as in \citep{DCAL,DMUX}. We describe this for the $1$-st fusion centre in the following, but the same calibration scheme is also performed by other fusion centres and their compute agents. The Jones matrices for the $k$-th direction, at frequency $f_i$, for all $N$ stations are represented in block form as
\beq
{{\bf J}_{kf_i}}\buildrel\triangle\over=[{\bf{J}}_{1k{f_i}}^T,{\bf{J}}_{2k{f_i}}^T,\ldots,{\bf{J}}_{Nk{f_i}}^T]^T,
\eeq
where ${{\bf J}_{k{f_i}}} \in \mathbb{C}^{2N\times 2}$. We represent the cost function (\ref{gLBFGS}) with ${\bf J}_{kf_i},\ k\in[1,K]$ as input by $g_{f_i}(\{{\bf J}_{kf_i}:\ \forall k\})$. It is straightforward to get $g({\bmath \theta})$ from $g_{f_i}(\{{\bf J}_{kf_i}:\ \forall k\})$ (and vice versa) by mapping ${\{{\bf J}_{kf_i}:\ \forall k\}}$ to ${\bmath \theta}$ and we omit the details here.

Following our previous work, we enforce smoothness in frequency by the constraint
\beq
{\bf J}_{kf_i}={\bf B}_{f_i} {\bf Z}_{k}^{(1)}
\eeq
where ${\bf B}_{f_i} \in \mathbb{R}^{2N\times 2FN}$ and ${\bf Z}_{k}^{(1)} \in \mathbb{C}^{2FN\times 2}$. The polynomial basis (with $F$ basis functions) evaluated at frequency $f_i$ is given by ${\bf B}_{f_i}$. The global variable (but {\em local} to the fusion centre $1$) is given by ${\bf Z}_{k}^{(1)}$. We remind the reader that we use the superscript $(\cdot)^{(1)}$ to denote that ${\bf Z}_{k}^{(1)}$ is local to fusion centre $1$ and its $C$ compute agents. Calibration with the frequency smoothness constraint is formulated as 
\beqn \label{conscalib}
\{{\bf {J}}_{kf_i},\ldots,{\bf {Z}}_k^{(1)}:\ \forall\ k,i\}=\underset{{\bf {J}}_{kf_i},\ldots,{\bf {Z}}_k^{(1)}}{\rm arg\ min} \sum_i g_{f_i}(\{{\bf J}_{kf_i}:\ \forall k\})\\\nonumber
{\rm subject\ to}\ \  {\bf {J}}_{kf_i}={\bf {B}}_{f_i} {\bf {Z}}_k^{(1)},\ \ i\in[1,P],k\in[1,K]\\\nonumber
{\rm and}\ \ {\bf {Z}}_k^{(1)}=\overline{{\bf Z}}_k, \ \ k\in[1,K].
\eeqn
The key difference from our previous work is the additional constraint ${\bf {Z}}_k^{(1)} = \overline{{\bf Z}}_k$, where $\overline{{\bf Z}}_k \in  \mathbb{C}^{2FN\times 2}$ is a global variable that is available to {\em all} fusion centres and this is calculated at the federated averaging centre in Fig. \ref{block}. {  We introduce Lagrange multipliers for each constraint, namely, ${\bf {Y}}_{kf_i}$ ($\in  \mathbb{C}^{2N\times 2}$) for the constraint ${\bf {J}}_{kf_i}={\bf {B}}_{f_i} {\bf {Z}}_k^{(1)}$, and, ${\bf X}_{k}$ ($\in \mathbb{C}^{2FN\times 2}$) for the constraint ${\bf {Z}}_k^{(1)}=\overline{{\bf Z}}_k$.} To find a solution for (\ref{conscalib}) at fusion centre $1$, { we need to minimize the original cost and the cost due to the constraints} and we form the augmented Lagrangian as 
\beqn \label{aug}
\lefteqn{
L(\{{\bf {J}}_{kf_i},{\bf {Z}}_k^{(1)},{\bf {Y}}_{kf_i},{\bf X}_{k}: \forall\ k,i\}) }\\\nonumber
&& =\sum_i g_{f_i}(\{{\bf J}_{kf_i}:\ \forall k\})\\\nonumber
&& + \sum_{i,k} \left( \| {\bf {Y}}_{kf_i}^H ({\bf {J}}_{kf_i}- {\bf {B}}_{f_i} {\bf {Z}}_k^{(1)})\| + \frac{\rho}{2} \| {\bf {J}}_{kf_i}- {\bf {B}}_{kf_i} {\bf {Z}}_k^{(1)} \|^2 \right) \\\nonumber
&&+\sum_k \left(\|{\bf X}_k^H\left(  {\bf {Z}}_k^{(1)} - \overline{{\bf Z}}_k \right)\|+ \frac{\alpha}{2} \|  {\bf {Z}}_k^{(1)} - \overline{{\bf Z}}_k \|^2 \right).
\eeqn
In (\ref{aug}), $\rho \in \mathbb{R}^{+}$ is the regularization factor for smoothness in frequency and $\alpha \in \mathbb{R}^{+}$ is the regularization factor for variable ${\bf {Z}}_k^{(1)}$ over all fusion centres. We use consensus alternating direction method of multipliers (ADMM) \citep{boyd2011} exactly as before \citep{DCAL,DMUX} to find solutions for ${\bf {J}}_{kf_i}$ and ${\bf {Y}}_{kf_i}$. The only difference is finding a solution for ${\bf X}_{k}$ and  ${\bf {Z}}_k^{(1)}$. The gradient of $L(\{{\bf {J}}_{kf_i},{\bf {Z}}_k^{(1)},{\bf {Y}}_{kf_i},{\bf X}_{k}: \forall\ k,i\})$ with respect to ${\bf {Z}}_k^{(1)}$ is 
\beqn \label{gradZ}
\lefteqn{
{2\times \rm grad}(L,{\bf {Z}}_k^{(1)})=}\\\nonumber
&& \sum_i {\bf {B}}_{f_i}^T\left(-{\bf {Y}}_{kf_i}+\rho\left(-{\bf {J}}_{kf_i}+{\bf {B}}_{f_i} {\bf {Z}}_k^{(1)}\right)\right)\\\nonumber
&& + {\bf X}_{k}+ \alpha \left({\bf {Z}}_k^{(1)} - \overline{{\bf Z}}_k \right)
\eeqn
and equating this to zero gives 
\beqn \label{zsol}
\lefteqn{
{\bf {Z}}_k^{(1)}= }\\\nonumber
&&  \left( \sum_i \rho {\bf {B}}_{f_i}^T {\bf {B}}_{f_i} + \alpha {\bf I}_{2FN} \right)^{\dagger} 
\left(\sum_i {\bf {B}}_{f_i}^T \left({\bf {Y}}_{kf_i} + \rho {\bf {J}}_{kf_i}\right) + \alpha  \overline{{\bf Z}}_k -{\bf X}_{k} \right).
\eeqn
The major difference in (\ref{zsol}) from our previous work is that the solution obtained for ${\bf {Z}}_k^{(1)}$ includes global information $\overline{{\bf Z}}_k$, which is directly fed into the solution by $\alpha\overline{{\bf Z}}_k$ and indirectly by the Lagrange multiplier ${\bf X}_{k}$. The global $\overline{{\bf Z}}_k$ is updated at regular intervals (not necessarily at each ADMM iteration) by federated averaging \citep{McMahan2016,Sava2019} -- i.e.,
\beq \label{fa}
\overline{{\bf Z}}_k = \frac{1}{D} \sum_{j=1}^{D} {\bf {Z}}_k^{(j)}
\eeq
where the $j$-th fusion centre sends ${\bf {Z}}_k^{(j)}$ to the federated averaging centre in Fig. \ref{block}. There is one caveat that we need to keep in mind when we find the average as in (\ref{fa}), when we have a sky model with unpolarized sources (which is the most common scenario). First, note that the solutions for (\ref{V}) -- i.e., ${\bf {J}}_{kf_i}$, can have a unitary ambiguity \citep{interpolation}. In other words, if ${\bf {J}}_{kf_i}$ is a valid solution, ${\bf {J}}_{kf_i} {\bf U}$ where ${\bf U} \in \mathbb{C}^{2\times 2}$ is an unknown unitary matrix, is also a valid solution. Therefore, if ${\bf {Z}}_k^{(j)}$ is a valid solution for (\ref{zsol}) at the $j$-th fusion center, then ${\bf {Z}}_k^{(j)} {\bf U}$ (where ${\bf U}\in \mathbb{C}^{2\times 2}$ is unitary) is also a valid solution. Hence, each ${\bf {Z}}_k^{(j)}$ in (\ref{fa}) will have its own unitary ambiguity and we use an iterative scheme as proposed by \cite{interpolation} to find the average $\overline{{\bf Z}}_k$. Moreover, fusion centre $j$ gets back $\overline{{\bf Z}}_k$ which is projected back to the current value of ${\bf {Z}}_k^{(j)}$ to minimize $\|\overline{{\bf Z}}_k {\bf U} - {\bf {Z}}_k^{(j)}\|$ where ${\bf U} \in \mathbb{C}^{2\times 2}$ is determined by solving the matrix Procrustes problem \citep{interpolation}. 

We make several remarks regarding (\ref{zsol}) here:
\begin{itemize}
\item The scheme of local consensus optimization together with global federated averaging is already being used in other applications, see e.g., \citep{Sava2019}, which we can use for further enhancement of our algorithm.
\item The averaging (\ref{fa}) assumes the number of datasets are equally distributed between each fusion centre and the associated compute agents. If this is not the case, a weighted average can be performed here. Moreover, in order to handle missing data or already flagged data, a similar weighting scheme can be applied.
\item Consider the case where the data available at each fusion centre span a narrow bandwidth, or in other words, the columns of ${\bf {B}}_{f_i}$ for all $f_i$ local to a fusion centre span a narrow region (compared to the case where the full set of frequencies is used to evaluate the basis). In this case, solving (\ref{zsol}) will be highly ill-conditioned. By having $\alpha>0$, we can reduce this ill-conditioning. In other words, via $\overline{{\bf Z}}_k$, we can feed information available at other frequencies (or other fusion centres) to each fusion centre. In the extreme case, by making $\alpha \rightarrow \infty$, we can force all fusion centres to only use the federated average as the solution for (\ref{zsol}).
\item Instead of using (\ref{fa}) for finding $\overline{{\bf Z}}_k$, we can use other sources of information as well. For instance, we can rely on physical models for the beam shape and the ionosphere \citep{DistModel,Albert2020} to derive $\overline{{\bf Z}}_k$ and feed this to calibration using (\ref{zsol}).
\item The update (\ref{fa}) does not have to be performed at the same cadence as the update of (\ref{zsol}). Moreover, if one fusion centre does not receive an updated value for $\overline{{\bf Z}}_k$, the calibration can be carried out by using and older value for $\overline{{\bf Z}}_k$ or, in the extreme case, by making $\alpha=0$. When $\alpha=0$, we revert to our previous calibration schemes \citep{DCAL,DMUX}. This is useful when the data are stored in multiple data processing clusters. Each fusion centre is connected to its compute nodes via a fast and reliable local network while the communication between each fusion centre and the federated averaging centre is through the slow and unreliable internet. Furthermore, we have to preserve privacy and security  when communicating via the internet and we can use a specialized communication scheme between the fusion centres and the federated averaging centre to achieve this.
\end{itemize}

We summarize the distributed stochastic calibration scheme in algorithm \ref{algSADMM}. There is basically three iterative loops in algorithm \ref{algSADMM}. We try to minimize the number of epochs $E$ since we read the data from disk. We also try to maximize $M$, the number of mini-bacthes, because the size of the data read into memory is proportional to $1/M$. We try to keep the maximum number of ADMM iterations $A$ as low as possible as well, to cut down the number of times data is read and also to cut down the network bandwidth use. We however caution that the best values of $A$, $E$ and $M$ need to be determined to suit each situation and it depends on many variables including the signal to noise ratio of the data, the number of constraints, the number of directions being calibrated $K$, the network bandwidth, and  the memory of each compute node as well as the disk reading speed.

\begin{algorithm}
\caption{Distributed stochastic calibration}
\label{algSADMM}
\begin{algorithmic}[1]
\REQUIRE Number of ADMM iterations $A$, Number of mini-batches $M$ and  Number of epochs $E$
\STATE Initialize ${\bf {Y}}_{kf_i}$,${\bf Z}_k^{(j)}$,${\bf X}_k$ to zero $\forall k,i,j$
\FOR{$a=1,\ldots,A$}
\STATE \COMMENT{In parallel at all compute agents and fusion centres}
\FOR{$e=1,\ldots,E$}
\FOR{$m=1,\ldots,M$}
 \STATE \COMMENT{Using $m$-th mini-batch of data, $\forall k,i,j$}
 \STATE Compute agents solve (\ref{aug}) for ${\bf {J}}_{kf_i}$
 \STATE Fusion centres solve (\ref{zsol}) 
 \STATE ${\bf {Y}}_{kf_i} \leftarrow {\bf {Y}}_{kf_i} + \rho \left({\bf {J}}_{kf_i}-{\bf {B}}_{f_i} {\bf {Z}}_k^{(j)}\right)$
\ENDFOR
\ENDFOR
\STATE Update federated average (\ref{fa})
\STATE ${\bf X}_k \leftarrow {\bf X}_k + \alpha \left(  {\bf {Z}}_k^{(j)} - \overline{{\bf Z}}_k\right)$
\ENDFOR
\end{algorithmic}
\end{algorithm}
In section \ref{sec:simul}, we test the performance of distributed stochastic calibration using simulated data. We also compare the performance of the stochastic LBFGS scheme with commonly used first order stochastic optimization algorithms.

\section{Simulations}\label{sec:simul}
The workhorse of our distributed stochastic  calibration scheme is the stochastic LBFGS algorithm presented in \citep{DSW2019,escience2018}. In contrast, there is a multitude of gradient descent based stochastic optimization methods that are widely used in machine learning \citep{robbins1951,Adam}. Therefore, we first compare the performance of the stochastic LBFGS algorithm with a widely used gradient descent based optimization method (Adam) \cite{Adam} in calibration of radio interferometric data. We have implemented the LBFGS algorithm in  PyTorch \citep{paszke} \footnote{https://github.com/SarodYatawatta/calibration-pytorch-test}, a popular machine learning software. 

We simulate an interferometric array with $N=62$ stations, observing a point source of $1$ Jy at the phase centre ($K=1$). The full batch size is $T=10$ times lots at a single frequency ($B=1$). We generate Jones matrices in (\ref{V}) with complex circular Gaussian entries having zero mean and unit variance. Finally, we add additive white Gaussian noise to the corrupted data with a signal to noise ratio of $0.1$. For calibration, we minimize the cost (\ref{minicost}) for this dataset with several mini-batch sizes $M$. The smallest mini-batch size is $M=1$ time slots and the largest is $M=10=T$, which also corresponds to full batch mode of calibration. The number of mini-batches for a single epoch is $T/M$ and varies from $10$ to $1$, respectively. For LBFGS, we use memory size of $7$ and $4$ iterations per each mini-batch. For Adam, we use a learning rate of $0.1$. We measure the performance in terms of the robust cost $g_i({\bmath \theta})$, evaluated for each mini-batch number $i$, and normalized by $1/M$. 

\begin{figure}
\begin{minipage}{0.98\linewidth}
\begin{minipage}{0.98\linewidth}
\centering
 \centerline{\epsfig{figure=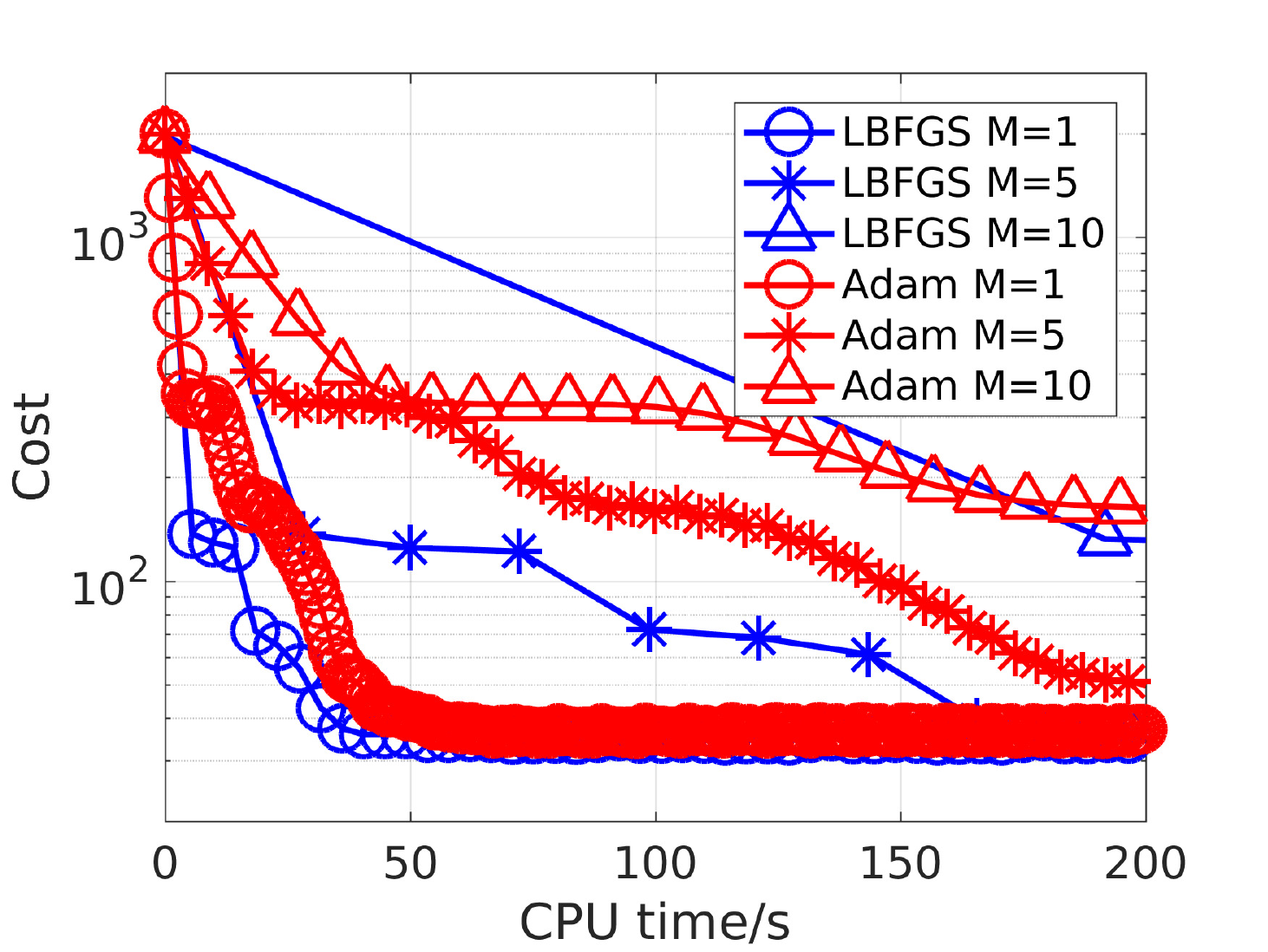,width=8.0cm}}
\vspace{0.1cm} \centerline{(a)}\smallskip
\end{minipage}\\
\begin{minipage}{0.98\linewidth}
\centering
 \centerline{\epsfig{figure=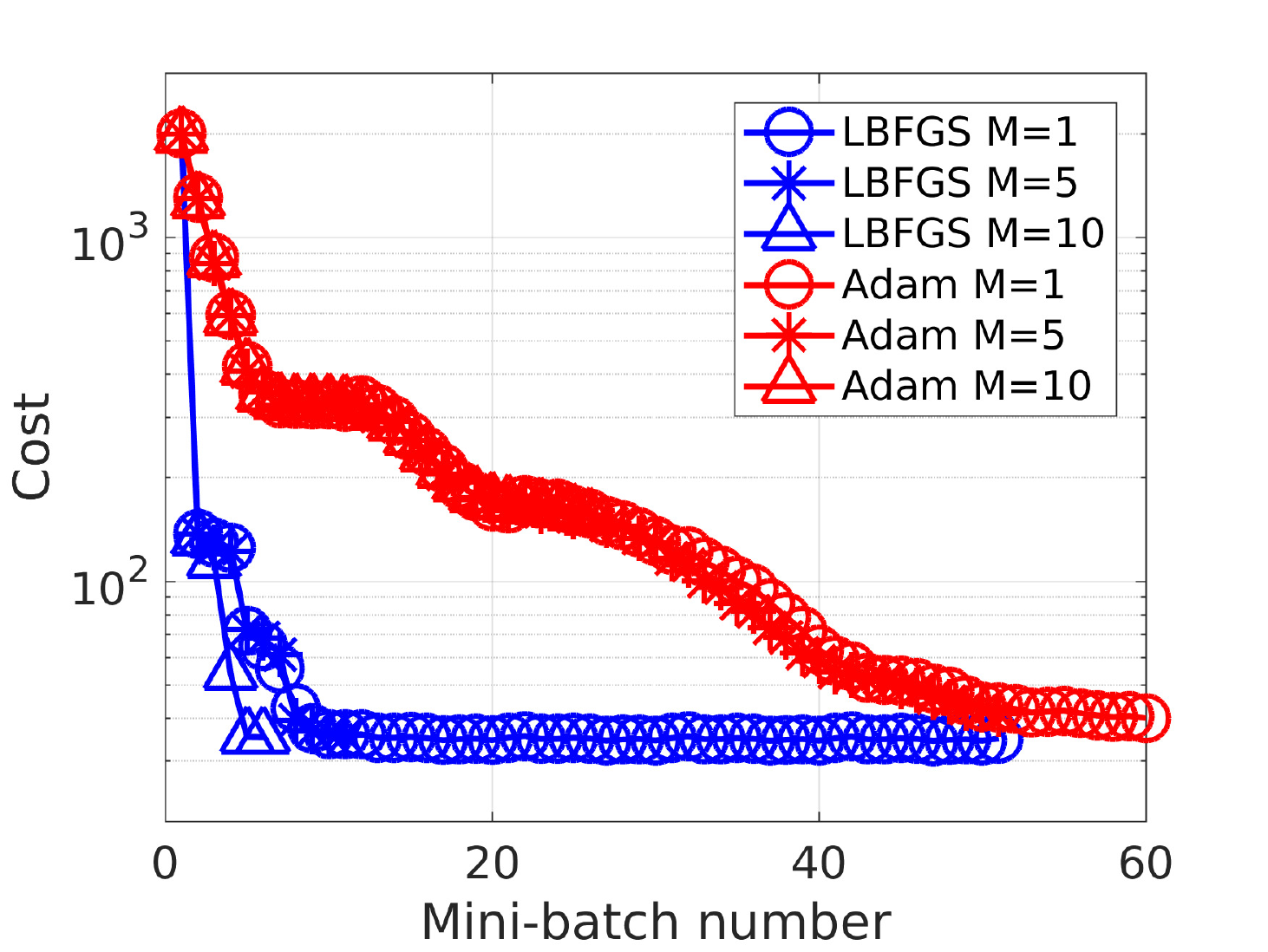,width=8.0cm}}
\vspace{0.1cm} \centerline{(b)}\smallskip
\end{minipage}
\end{minipage}
\caption{Comparison of LBFGS with Adam for various mini-batch sizes $M$. (a) The reduction of cost with compute time. (b) The reduction of cost with mini-batch number. While Adam uses much less compute time to calibrate each mini-batch, it convergences much slower. \label{adam_comp}}
\end{figure}

We show the results of the comparison between LBFGS and Adam in Fig. \ref{adam_comp}. In Fig. \ref{adam_comp} (a), we plot the reduction of the cost with computing time (measured using a single CPU) while in Fig. \ref{adam_comp} (b), we show the reduction of the cost with each mini-batch of data processed. We see that Adam runs much faster than LBFGS, processing more mini-batches of data within a given CPU time interval. However, the convergence (or the reduction of cost) of Adam is slower than LBFGS, illustrating their first-order and second-order convergence rates which is well known \cite{Fletcher,Liu1989}. In particular, if we count the number of mini-batches required for each algorithm to reach convergence, we see in Fig. \ref{adam_comp} (b) that LBFGS needs far fewer mini-batches. As shown in Fig. \ref{block}, we need to read each mini-batch of data from disk, and the cost of reading data is much less for LBFGS, making it the preferred choice for stochastic calibration. 

Having established the superiority of stochastic LBFGS for our particular use case, we test the performance of distributed stochastic calibration in the next simulation. In order to do this, we again simulate an inteferometric array with $N=62$ stations (similar to LOFAR \citep{LOFAR}). Data is simulated over $8$ subbands uniformly distributed in the frequency range $115$ MHz to $185$ MHz, and each subband has $64$ channels each, with each channel having a bandwidth of $3$ kHz. The bandwidth of each subband is $0.192$ MHz and the total bandwidth is therefore $1.536$ MHz. The total observation time is $20$ minutes, with data sampling at every $1$ second. Therefore, the total number of datapoints is $1200$. 

\begin{figure}
\begin{minipage}{0.98\linewidth}
\begin{minipage}{0.98\linewidth}
\centering
 \centerline{\epsfig{figure=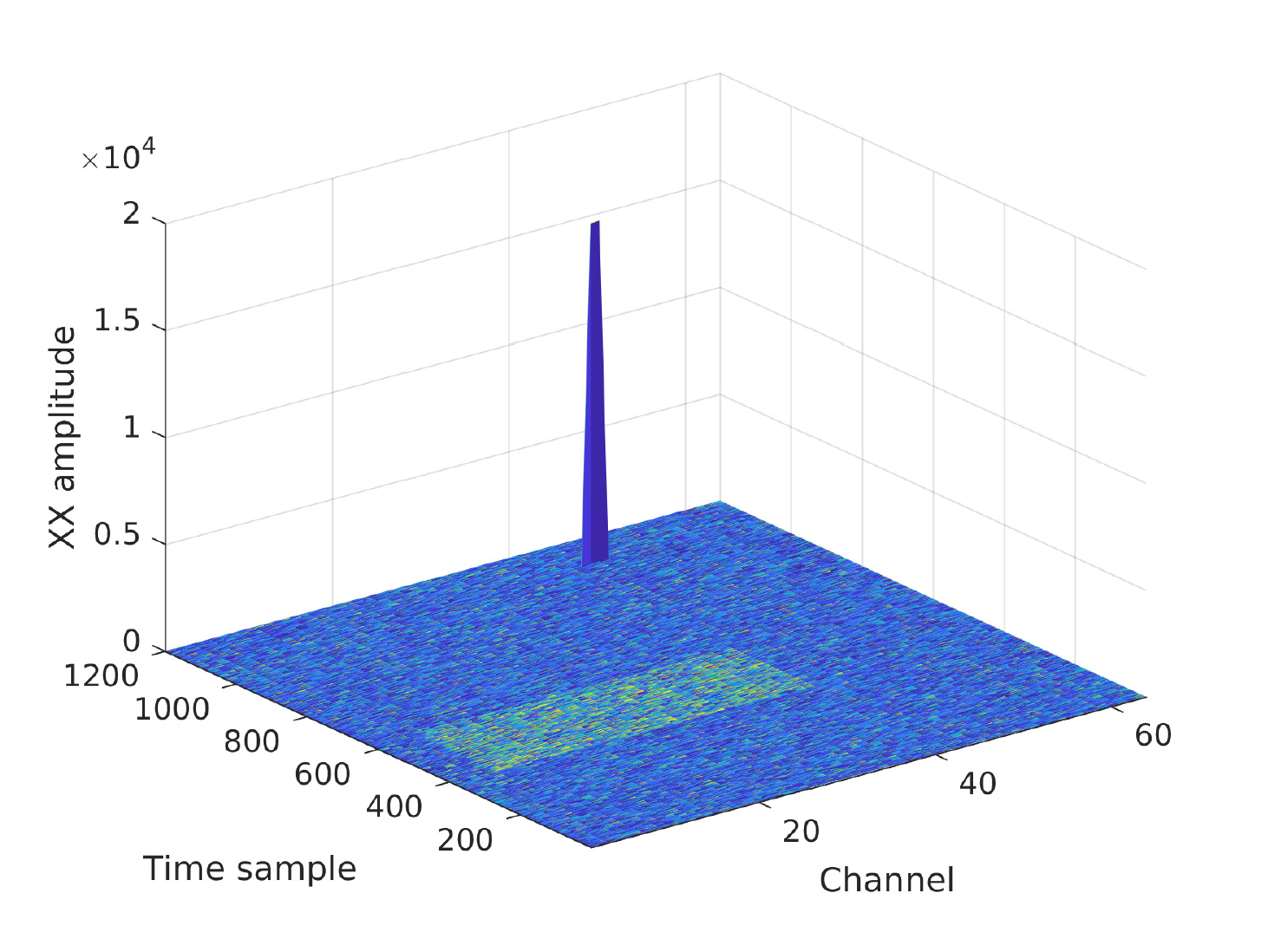,width=8.0cm}}
\vspace{0.1cm} \centerline{(a)}\smallskip
\end{minipage}\\
\begin{minipage}{0.98\linewidth}
\centering
 \centerline{\epsfig{figure=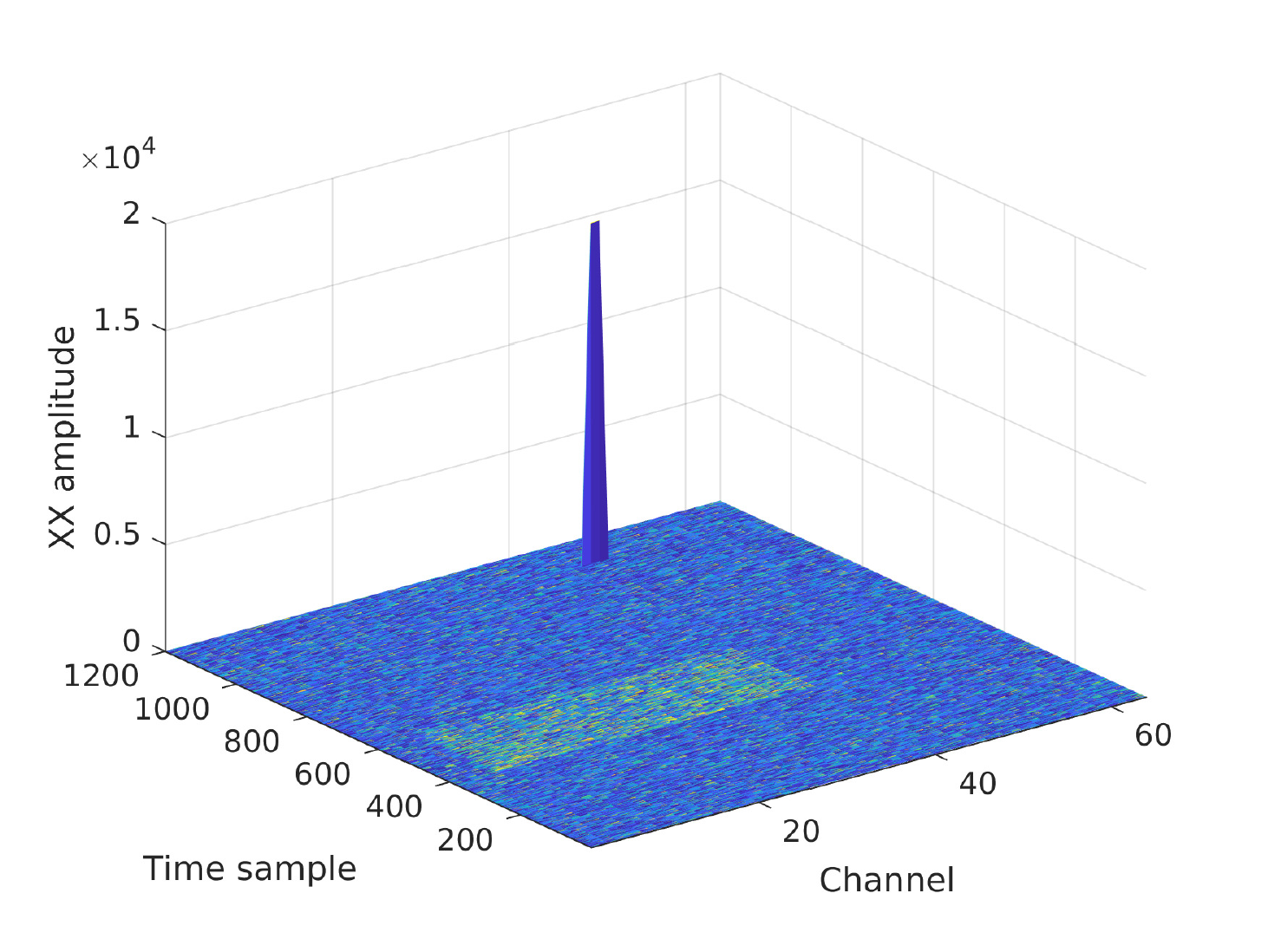,width=8.0cm}}
\vspace{0.1cm} \centerline{(b)}\smallskip
\end{minipage}
\end{minipage}
\caption{Visibility amplitude of $XX$ correlation of one baseline, showing both narrow band high and broad band low RFI (a) before calibration, and (b) after calibration. \label{XXamp}}
\end{figure}

We simulate $K=2$ points sources in the sky (with flux densities $3$ Jy and $1.5$ Jy) and corrupt their signals with direction dependent systematic errors. The systematic errors are modeled as Jones matrices with complex circular Gaussian entries with zero mean and unit variance. Moreover, the systematic errors are randomly varied for every $10$ time samples and also varied smoothly over frequency (by multiplying them with low order polynomials in frequency). An additional $150$ weak sources (flux density in the range $0.01$ Jy to $0.1$ Jy) randomly positioned across a field of view of $7\times 7$ square degrees are also added to the simulation, but without any systematic errors (mainly to check the accuracy of calibration). Next, the total signal is multiplied by a random and smooth bandpass polynomial, per each subband. Additive white Gaussian noise with a signal to noise ratio of $0.1$ is added to this signal. Finally, Radio frequency interference is also added. Both broad band (long duration), low amplitude as well as narrow band (narrow duration), high amplitude RFI is randomly simulated and added for a randomly selected subset of baselines. We show a typical sample of RFI added to the data in Fig. \ref{XXamp} (a).

\begin{figure}
\begin{minipage}{0.98\linewidth}
\centering
 \centerline{\epsfig{figure=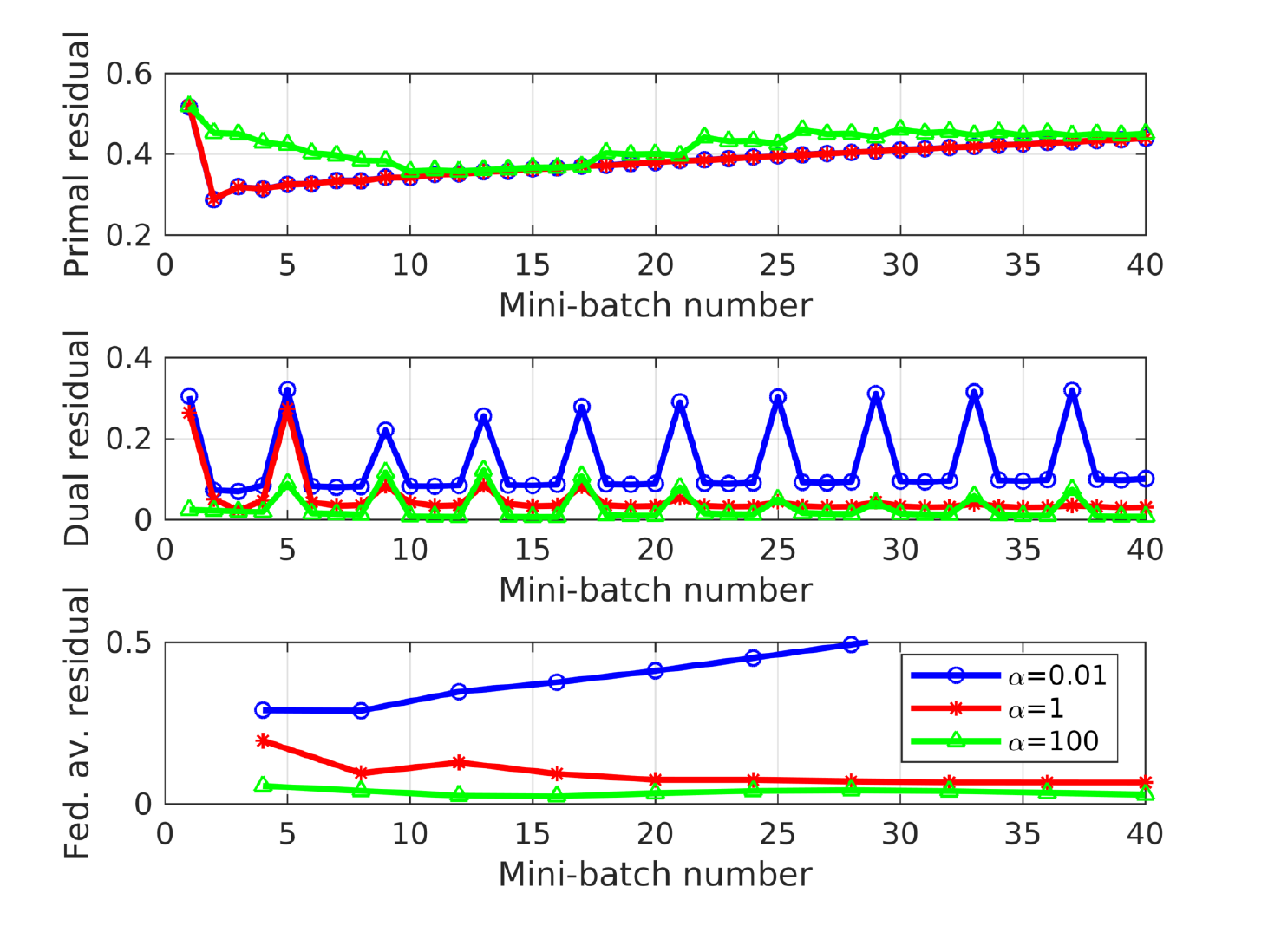,width=8.0cm}}
\end{minipage}
\caption{Convergence for three different values of $\alpha$. We show the dual residual, primal residual and federated averaging residual from top to bottom.\label{convg}}
\end{figure}

For consensus optimization, we construct a Bernstein basis with $F=3$ basis functions. The basis spans the full frequency range  $[115,185]$ MHz. In order to calibrate this dataset, we use $D=8$ fusion centres (see Fig. \ref{block}) and each fusion centre works with $C=P=16$ compute agents. The original $64$ channels of each subband is averaged to $B=16$ for obtaining a solution. We obtain a solution for every $T=10$ time samples using a mini-batch size of $M=5$ (so $2$ mini-batches in total). We show the performance of calibration in terms of three criteria in Fig. \ref{convg}. We measure the primal residual $\| {\bf {J}}_{kf_i}- {\bf {B}}_{kf_i} {\bf {Z}}_k^{(j)} \|$, the dual residual $\| \left({\bf {Z}}_k^{(j)}\right)^{new} - \left({\bf {Z}}_k^{(j)}\right)^{old} \|$ and the federated averaging residual $\| {\bf {Z}}_k^{(j)} - \overline{{\bf Z}}_k \|$ with each iteration (or mini-batch). We have normalized each quantity in Fig. \ref{convg} by the sizes of the matrices involved and find the average value for all $k,i,j$. We have $E=2$ and $M=2$ and because of this, the federated averaging residual is updated at every $E\times M=4$ mini-batches. We have varied the regularization factor for federated averaging -- i.e., $\alpha=0.01,1,100$ while keeping the consensus penalty at $\rho=1.0$.

We make several observations from Fig. \ref{convg}. First, we see that the primal residual shows no improvement with iterations, this is because our basis functions with $F=3$ do not have enough freedom to completely describe the frequency behavior of the systematic errors. We remind the reader that in addition to the global variation within the full band of $[115,185]$ MHz, there is local variation within $0.192$ MHz of each subband due to the bandpass shapes we have introduced. Therefore, the systematic errors have more variation than what is assumed by the consensus polynomials and the primal residual reflect this. The marked difference in performance is seen in the dual residual and the federated averaging residual. For $\alpha=0.01$, the regularization is too low for the federated averaging to come into effect and the federated averaging residual diverges. In other words, each fusion centre finds a solution for ${\bf {Z}}_k^{(j)}$ that is much different from the federated average. When $\alpha=100$, the federated averaging is forced upon each fusion centre and this gives poor performance as seen in the dual and primal residuals. The best result is obtained at $\alpha=1$, when we see both dual residual and federated averaging residual go to a low value together and in this case, we can say that each fusion centre has a solution ${\bf {Z}}_k^{(j)}$ that is also globally accepted. This is what we want to achieve in terms of the physical origins of the systematic errors.

\begin{figure}
\begin{minipage}{0.98\linewidth}
\begin{minipage}{0.98\linewidth}
\centering
 \centerline{\epsfig{figure=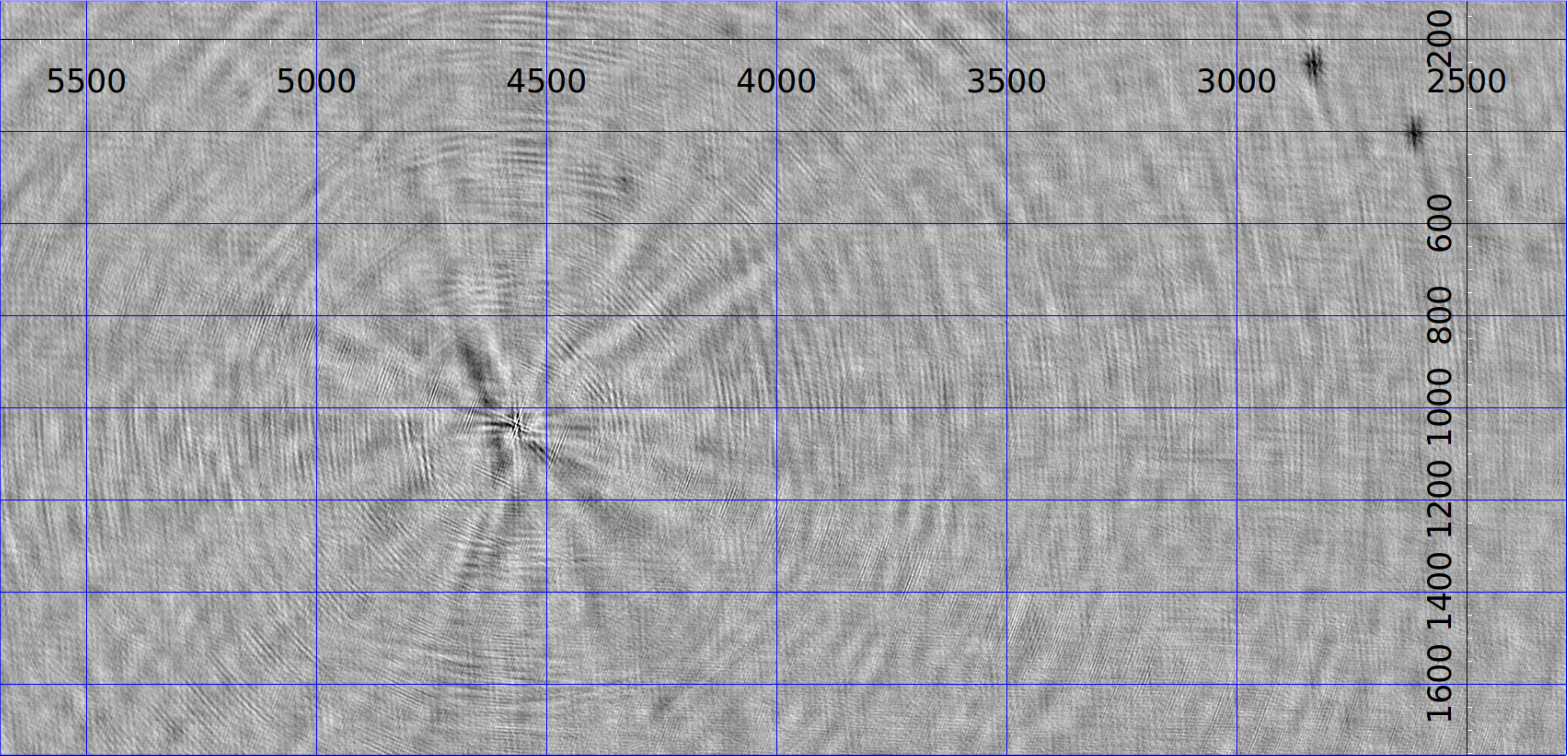,width=8.0cm}}
\vspace{0.1cm} \centerline{(a)}\smallskip
\end{minipage}\\
\begin{minipage}{0.98\linewidth}
\centering
 \centerline{\epsfig{figure=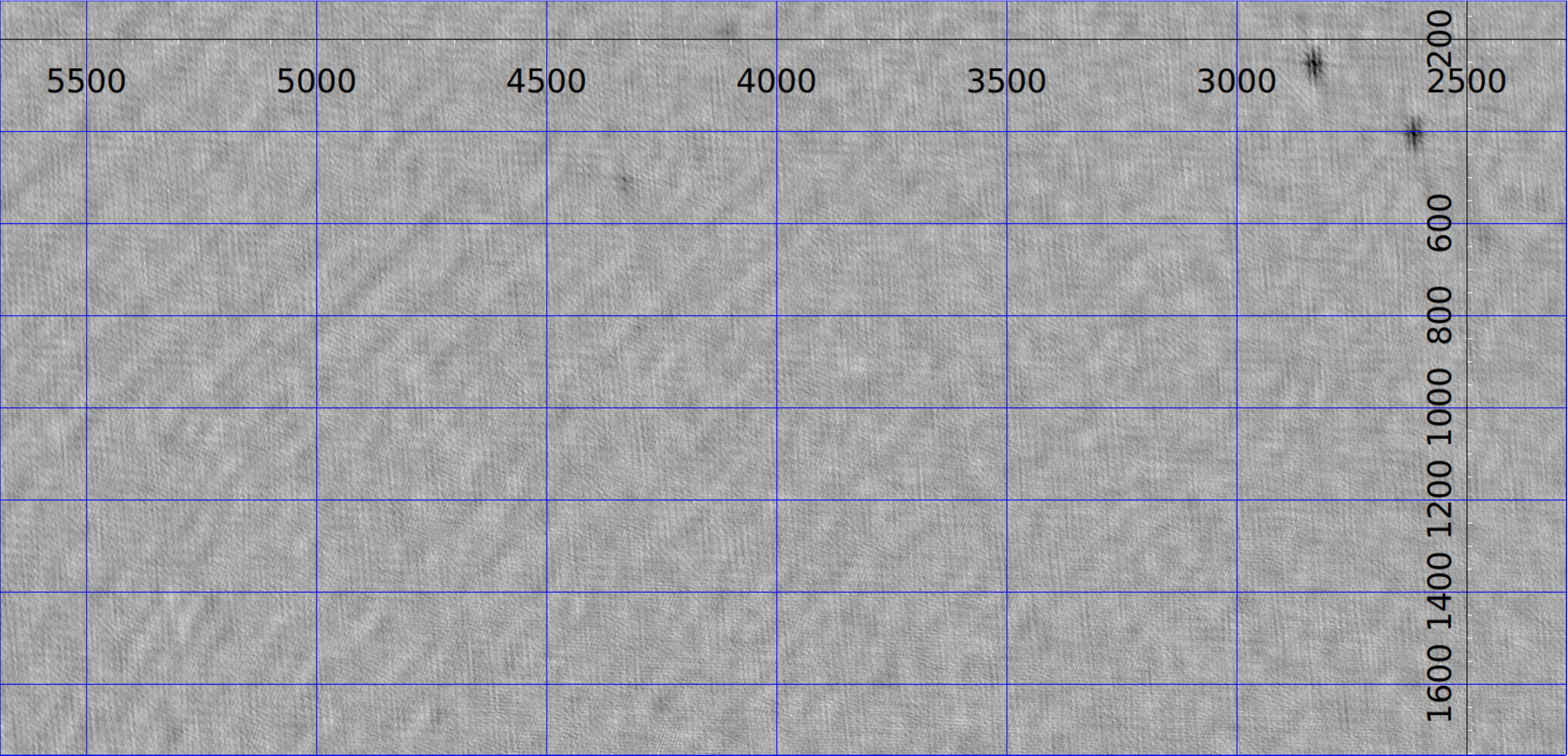,width=8.0cm}}
\vspace{0.1cm} \centerline{(b)}\smallskip
\end{minipage}
\end{minipage}
\caption{Images showing the area around the strongest source $3$ Jy peak flux (a) before calibration, and (b) after calibration. Weak sources and artefacts due to RFI are still present after calibration. \label{sI}}
\end{figure}

We calibrate the full dataset with the number of ADMM iterations set to  $A=4$ and with $\alpha=1$. The total number of mini-batches used for each calibration run is therefore $A\times E\times M=4\times 2\times 2=16$ and this is lower than in Fig. \ref{convg}. The solutions are initialized with the solution obtained for the averaged data of the first subband for the first $10$ time samples. We show the images of a small area in the sky surrounding the brightest source being calibrated ( $3$ Jy ) in Fig. \ref{sI}. We see that the contribution from this source is cleanly removed from the data and in Fig. \ref{sI} (b), only the weak sources and the RFI remain (see Fig. \ref{XXamp} (b)). This sky-subtracted data can be used for better RFI mitigation as in \citep{Wilensky2019}. Considering the computational effort, we use $D=8$ fusion centres, each working with a subband of data. Therefore, the network traffic at the federated averaging centre has to deal with $8$ messages at a time. In contrast, in our previous distributed calibration software, either we need to use $D\times P=8 \times 16=128$ fusion centres \citep{DCAL} (high network traffic) or we need to multiplex data by a factor of $1/16$ \citep{DMUX} (slow convergence).

Looking back at Fig. \ref{XXamp}, we see that the RFI is well preserved even after calibration. This is attributed to the robust cost function used in calibration \cite{Kaz3} and this is also confirmed by \citep{sob2019}. Therefore, after stochastic calibration, better RFI mitigation can be achieved using techniques similar to \citep{Wilensky2019}.

\section{Conclusions}\label{sec:conc}
We have presented a distributed stochastic calibration scheme that minimizes the use of compute memory and network traffic for the calibration of large data volumes at their highest resolution. We have also highlighted the many applications of this calibration scheme in radio astronomy. Ready to use software based on this scheme is already available\footnote{http://sagecal.sourceforge.net/} and we will explore this further for various science cases in radio astronomy in future work.

\section*{Acknowledgments}
This work is supported by Netherlands eScience Center (project DIRAC, grant 27016G05). We thank the anonymous reviewer for the valuable comments.

\bibliographystyle{mnras}
\bibliography{references}
\bsp
\label{lastpage}
\end{document}